\documentclass[11pt,twoside]{article}
\usepackage{epsfig}
\usepackage{here}
\usepackage{amsmath}
\usepackage{graphics}
\usepackage{subfigure}
\usepackage{cite}
\usepackage{hhline}

\newcommand{\reaktion}{\mbox{$pp\rightarrow\,pp\:\!K^+K^-$ }}
\newcommand{\reaktionbf}{$\boldsymbol{pp\rightarrow\,pp\:\!K^+K^-\ }$}

\begin{document}

\begin{center}{\bf\Large  Towards the excitation function of the \reaktionbf reaction at COSY-11}
\end{center}
\vspace{0.5cm}
\begin{center}
   P. Winter$^{a}$ for the COSY-11 collaboration\\[0.1cm]
   {\small \em 
         $^a$Institut f\"ur Kernphysik, Forschungszentrum J\"ulich, Germany
   }
\end{center}
\vspace{0.5cm}
\begin{center}
 \parbox{0.9\textwidth}{
  \small{
    {\bf Abstract:}\
The ongoing discussion concerning the nature of the scalar resonances $f_0(980)$ and $a_0(980)$ implicated to extend the studies of the $pp\rightarrow pp K^+K^-$ reaction near the production threshold. Furthermore, such elementary production processes allow to study the interaction of the outgoing particles due to their low relative momenta. Therefore, final state interactions in the $pK$ or $K\bar{K}$ system can be perfectly studied in such experiments.\par
The acquired data at the excess energies $Q=10$ and $28\,$MeV have been taken at COSY-11 and the status of the ongoing analysis will be presented.}}
\end{center}

\vspace{0.5cm}
\section{Introduction}
\begin{figure}[H]
\parbox{0.4\textwidth}
{\rotatebox{-90}{\epsfig{file=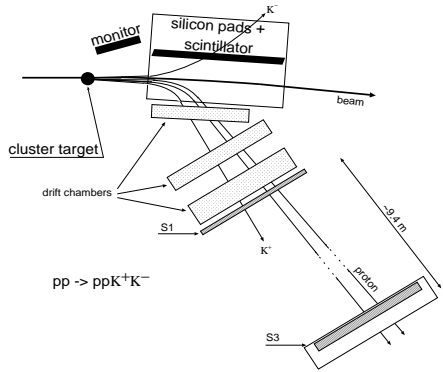,width=0.39\textwidth}}}
\hfill
\parbox{0.5\textwidth}
{\caption{\label{cosy11} \small Schematic top view of the COSY-11 detection setup with only the relevant parts for the measurement of the reaction \reaktion\hspace{-0.98ex}. The different components are described in the text.}}
\end{figure}

The experimental facility used for the studies described below is the internal COSY-11 detection setup \cite{brauksiepe:96} at the COoler SYnchrotron COSY \cite{maier:97nim}. Its schematical assembly with the relevant detector parts is shown in figure \ref{cosy11}.

The hydrogen cluster target is mounted in front of one of the COSY bending dipole magnets which acts like a magnetic spectrometer for the outgoing interacting particles. The positively charged particles are bend towards the interior of the ring and are then detected by a set of 3 drift chambers allowing for a momentum reconstruction. The following time of flight (TOF) measurement -- with two scintillator hodoscopes S1 and S3 -- in combination with the momentum reconstruction gives access to the particles' four momenta. The remaining negative kaon in the reaction \reaktion is registered in a detector mounted in the inner part of the dipole gap consisting of a scintillator and granulated silicon pads. Since the four momentum of the $K^-$ cannot be directly deduced, it is identified by means of the missing mass method.\\
In order to calculate an absolute cross section from the extracted number of events one needs a normalisation in form of the luminosity. Therefore, in this proton proton collision, a trigger was set up to record in parallel the elastic scattered protons. While one proton is registered in the main detector part, the second has to pass the so called monitor detector, a combination of a scintillator and a silicon pad detector.

\section{Motivation}
\begin{figure}[h]
\parbox{0.55\textwidth}
{\epsfig{file=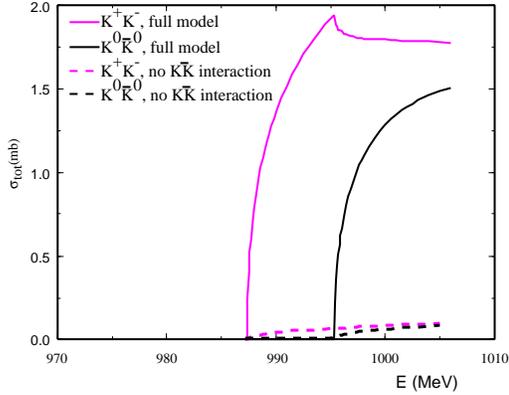,width=0.55\textwidth}}
\hfill
\parbox{0.4\textwidth}
{\caption{\label{krehl} \small Excitation function of the $\pi^+\pi^-\to K\bar{K}$ \cite{krehl:97} calculated in a one boson exchange model. Inclusion of the $K\bar{K}$ interaction (solid lines) strongly rises the slope of the total cross section for both the charged and uncharged channel.}}
\end{figure}

Elementary meson production in nucleon nucleon collisions near the production threshold enables to study besides the production mechanism also the interactions among the outgoing particles. Especially for short living particles, these studies are only possible in such production reactions. Therefore, the studies of the excitation function of the reaction \reaktion will help to understand the $K\bar{K}$ interaction as well as possible final state effects in the $pK$ system.\\
Calculations of the $\pi^+\pi^-\to K\bar{K}$ \cite{krehl:97} based on the J\"ulich model \cite{janssen:95} show a strong difference in the total cross section when switching on and off the $K\bar{K}$ interaction (see figure \ref{krehl}). The strong rise with inclusion of the $K\bar{K}$ interaction (solid lines) leads to the interpretation of the scalar meson $f_0(980)$ as a $K\bar{K}$ molecule.

\begin{figure}[h]
\parbox{0.55\textwidth}
{\epsfig{file=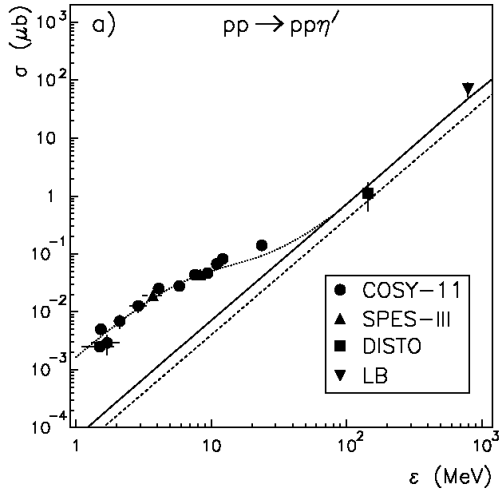,width=0.55\textwidth}}
\hfill
\parbox{0.4\textwidth}
{\caption{\label{ppetaprime} \small Cross section data for the reaction $pp \to pp \eta'$ \cite{hibou:98, moskal:98-01, moskal:00, balestra:00-02, khoukaz:04}. The dashed line represents an arbitrarily normalized phase space integral and the solid a calculation in a meson exchange model without inclusion of the FSI \cite{sibirtsev:97}. The dotted line incorporates the $pp$-FSI by including the on-shell $pp$ amplitude in order to reproduce the phase space population.}}
\end{figure}

Up to now, various scenarios have been proposed to describe the still unknown structure of the $f_0(980)$ meson, namely -- besides the mentioned $K\bar{K}$ molecule \cite{weinstein:90, lohse:90} -- a usual $q\bar{q}$ state \cite{morgan:93} or a $qq\bar{q}\bar{q}$ configuration \cite{jaffe:77}. Furthermore, other scenarios forsee a hybrid $q\bar{q}$/meson-meson sytem due to the strong coupling to S-wave two-meson channels \cite{beveren:86} or even quarkless gluonic hadrons \cite{jaffe:75}. Since the analysis of the reaction \reaktion gives a quite model independent way to learn about the $K\bar{K}$ interaction strength this might shed light on the open question concerning the $f_0(980)$ meson.\\
Another issue linked with the near threshold meson production is the fact that the outgoing particles have low relative momenta and hence, possible final state interaction (FSI) effects are pronounced. A prominent example is the $pp$-FSI which shows up for example in $pp \to pp\eta'$ at excess energies below $\sim$\,40\,MeV (compare figure \ref{ppetaprime}), where the cross section no longer follows a pure phase space distribution or the calculations without the $pp$-FSI (compare the pure $Q^2$ phase space dependenece (dashed line) and a calculation without FSI \cite{sibirtsev:97}). Only the dotted line reproduces the set of data at low energies well since it accounts for the proton-proton FSI.

This example shows that such final state effects are pronounced at low excess energies. Therefore, further data of the reaction \reaktion will help to investigate the strength of the interaction in the $pK$ system.

\section{Results} 
\begin{figure}[h]
\subfigure[\label{missmass-a}]{\epsfig{file=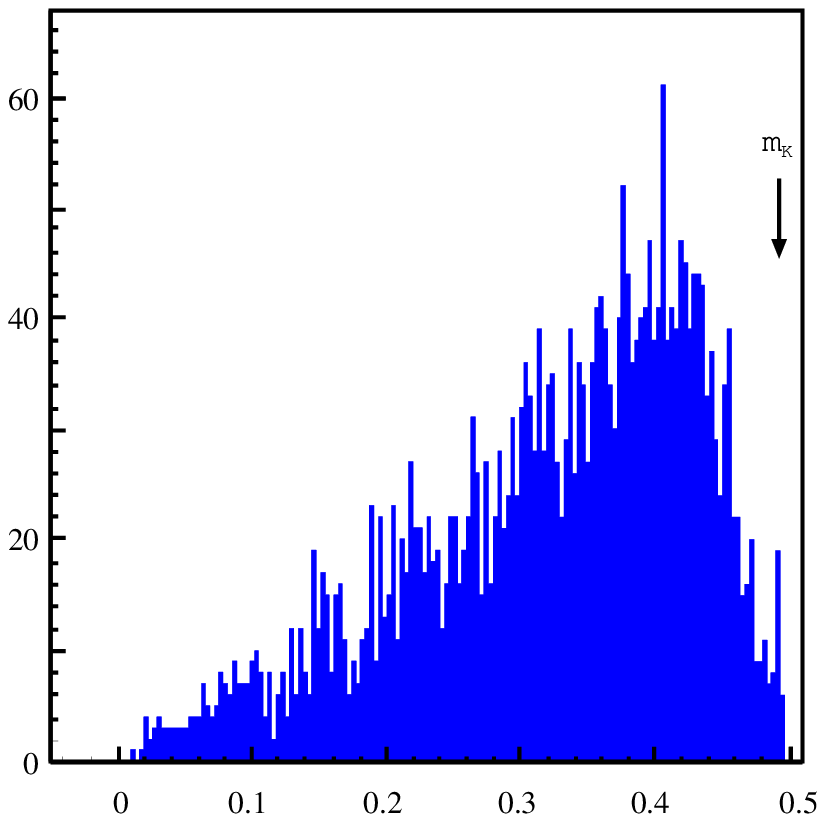,width=0.4\textwidth}}
\hfill
\subfigure[\label{missmass-b}]{\epsfig{file=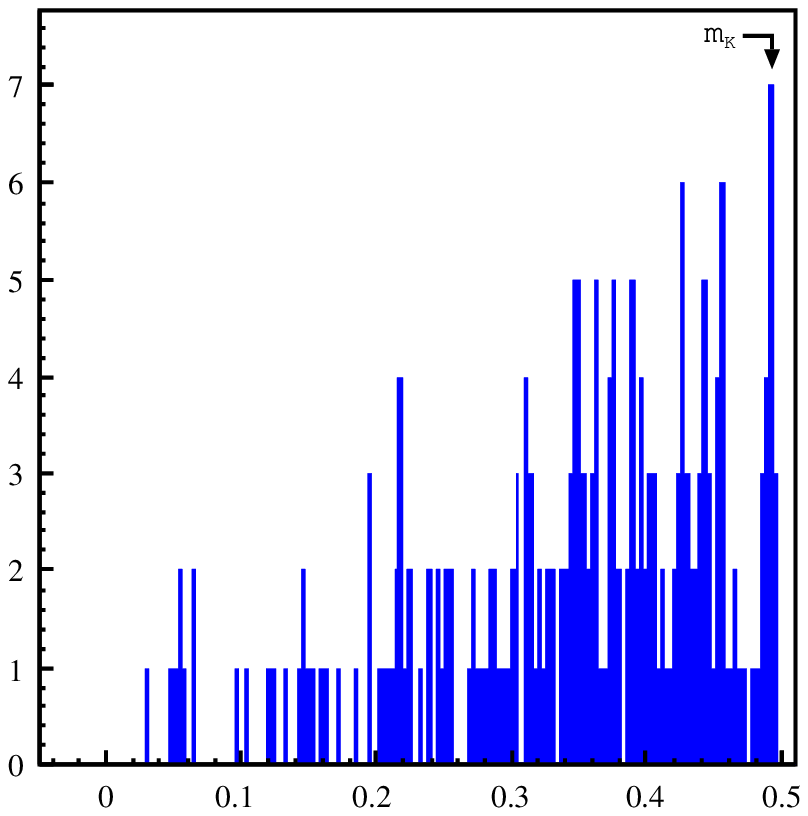,width=0.4\textwidth}}
\caption{\small Missing mass for the identified system $ppK^+$ for a) all events and b) with an additional request for a hit in the scintillator mounted in the dipole gap.}
\end{figure}

The COSY-11 collaboration has extended its previous measurement at an excess energy of $Q=17\,$MeV \cite{quentmeier:01-2} at two new energies, namely $Q=10$ and 28\,MeV, both lying below the $\phi$ threshold. The analysis is based on the identification of two protons in the exit channel via their mass. They are registered in the drift chambers and by tracing back the trajectory through the known magnetic field, their momentum is deduced. The subsequent TOF measurement between the two scintillators S1 and S3 over a distance of about 9.4\,m then gives in combination with the momentum the invariant mass, i.e. the four momentum. Due to their decay, the number of kaons registered in the S3 scintillator is rather low. Therfore, the S1 scintillator is used as the stop counter for the TOF measurement. The start time is the time of the interaction itself which is calculable via the momentum of the identified protons \cite{wolke:97}. The remaining negative kaon is then identified via the missing mass of the $ppK^+$ system, shown in figure \ref{missmass-a} for the data set at $Q=10$\,MeV. Besides the physical background mainly stemming from the reactions $pp \to pK^+ \Lambda(1405) / \Sigma(1385)$ and from misidentified pions \cite{quentmeier:01-2} a signal at the $K^-$ mass is visible.  The additional demand for a hit in the scintillator mounted in the dipole gap reduces much more the background than the signal (see figure \ref{missmass-b}). Similar spectra were also obtained for the higher energy at $Q$=28\,MeV where the $K^-$ peak is apparent.

\begin{figure}[h]
\centerline{
\subfigure[\label{korrelation_mc}]{\epsfig{file=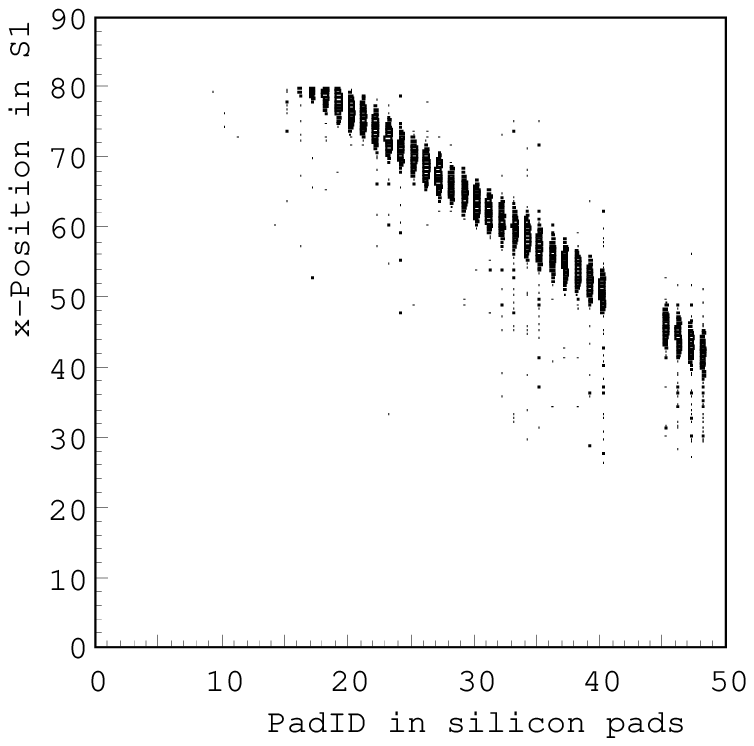,width=0.4\textwidth}}
\hspace{0.5cm}
\subfigure[\label{korrelation_data}]{\epsfig{file=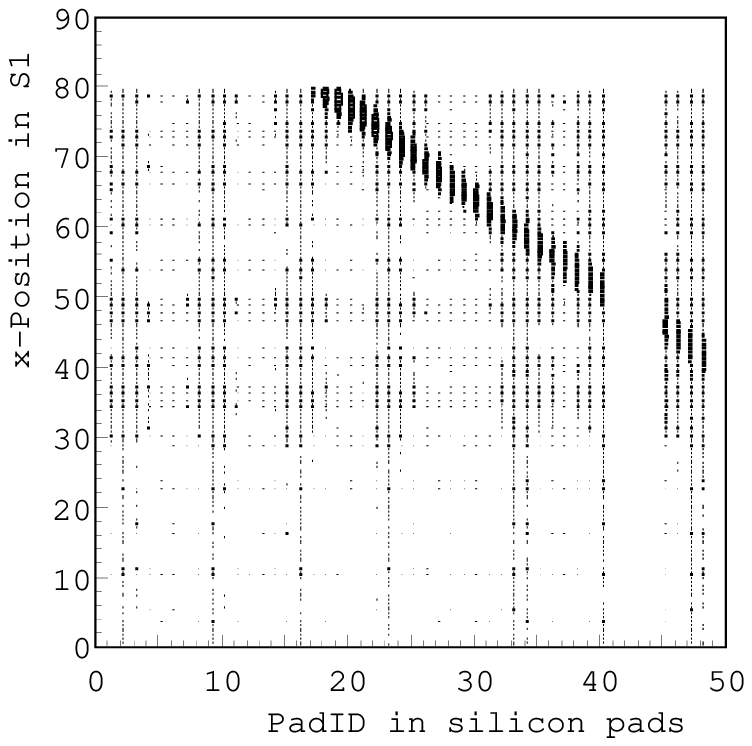,width=0.4\textwidth}}}
\caption{\small x-Position of the first particle in the S1 detector versus the pad number in the silicon pads behind the monitor scintillator for a) Monte Carlo simulations and b) for real data. Taken into account are events with one reconstructed track and a simultaneous hit in the monitor scintillator. The pads with number 41 to 44 were excluded from the analysis due to the fact, that they were not working properly during the experiment.\label{korrelation}}
\end{figure}

For the final cross sections, a further step is the determination of the luminosity via the reaction $pp \to pp$ recorded in parallel. While it is required that one proton is registered in the drift chambers and in the S1 scintillator, the other one has to pass the monitor detector which is installed towards the outer part of the accelerator in front of the dipole magnet. Due to the two body kinematics, there is a fixed correlation between the position of the first proton in the S1 detector and the position of the other proton passing the monitor detector. Figure \ref{korrelation} shows the x-position of the track in the S1 detector versus the pad number corresponding to the spatial crossing position of the particle for events with one reconstructed track and a hit in the monitor scintillator. The clear correlation band expected from Monte Carlo simulation (see fig. \ref{korrelation_mc}) is clearly visible in the data as well (compare fig. \ref{korrelation_data}) with only a low background outside of the correlation area mainly from the reaction $pp \to pn\pi^+$. 

By cutting out the correlation region, an extraction of the elastically scattered events is possible. Details of the whole standard procedure can be found elsewhere \cite{wolke:97, moskal:98} and for this specific analysis soon in ref. \cite{winter:05}. The analysis includes several different steps using Monte Carlo studies for the efficiency determination in case of the elastic reaction $pp \to pp$. Up to now, the total integrated luminosity $\mathcal{L}$ for both energies were extracted to be:\\
\begin{center}
\begin{tabular}{|c|c|}
\hline
\bf Q\ [MeV] & $\boldsymbol{\mathcal{L}\ [cm^{-2}]}$ \\
\hhline{|:=|=:|}
10\,MeV & $2.71 \cdot 10^{36}$\\
\hline
28\,MeV & $2.31 \cdot 10^{36}$\\
\hline
\end{tabular}
\end{center}

The statistical and systematical errors were not yet deduced from the data but will be available soon.

\section{Outlook}
The new data the COSY-11 collaboration acquired in \reaktion are still under analysis. The total integrated luminosity has been determined whereas systematic error studies are still in progress. Missing mass spectra of the $ppK^+$ system show a clear signal at the $K^-$ mass. Prior to the determination of the absolute cross sections, there will be a further cut performed in order to get the spectrum nearly background free. Since the four momentum of the $K^-$ is calculable, the measured position of the hit in the dipole scintillator can be compared with the expected position in this detector by tracking the $K^-$ momentum through the magnetic field. From the former analysis \cite{quentmeier:01-2}, it is known that this crucial last step eliminates the background and therefore will be applied for the new data, too. In combination with efficiency studies via Monte Carlo simulations, the number of registered events will give the absolute cross sections at $Q=10$ and 28\,MeV.

%\bibliography{abbrev,general}

\end{document}